\documentclass{kluwer}    

\usepackage{epsfig}   
\begin{document}
                 
\begin{article}
\begin{opening}         
\title{Mass loss from galaxies: feeding the IGM, recycling in the IGM} 
\author{P.-A. \surname{Duc}$^{*}$,
J. \surname{Braine}, 
U.  \surname{Lisenfeld},
P.  \surname{Amram} and 
E. \surname{Brinks} }  
\runningauthor{P.--A. Duc et al.}
\runningtitle{Mass loss from galaxies}
\institute{$^{*}$CEA Saclay, Service d'astrophysique,
 91191 Gif sur Yvette cedex, France}

\begin{abstract}
As a result of internal processes or
environmental effects like ram-pressure stripping or collisions, galaxies
 lose a significant part of their stellar and gaseous content.
Whereas the impact of such stripping on  galaxy evolution has been
well studied, much less attention has been given to the fate of the
 expelled material in the intergalactic or intracluster medium (IGM/ICM).
Observational evidence exists showing that a fraction of the injected 
  matter is actually recycled to form a new generation of galaxies,
 such as the Tidal Dwarf Galaxies discovered near numerous interacting
 systems. Using a set of multiwavelength data, we are now
able to roughly analyze the processes pertaining to their formation: from an
instability in the HI clouds, through the formation of molecular gas, and to
the onset of star formation.
\end{abstract}

\end{opening}           

\section{Loss of galactic material during galaxy evolution}  
Galaxy evolution goes hand in hand  with the loss of interstellar matter.
Many processes, of internal or external origin, contribute
to strip galaxies from their raw material.  
Starbursts and associated superwinds or active galactic nuclei via
 jets cause the ejection of plasmoids at distances of up to ten
 kpc. Such mechanisms do not involve large
quantities of matter but play a major role in enriching
the IGM/ICM with heavy elements and at the same
time in  regulating the chemical evolution of galaxies.
External processes have an even more dramatic effect on galaxy
 evolution. Whereas in clusters ram-pressure exerted by the ICM is
 efficient at stripping gaseous material, tidal forces act both on
 stars and gas, pulling them out  up to distances of 100 kpc.
Figure~1 illustrates several of these mechanisms.

\begin{figure} 
\centerline{\psfig{file=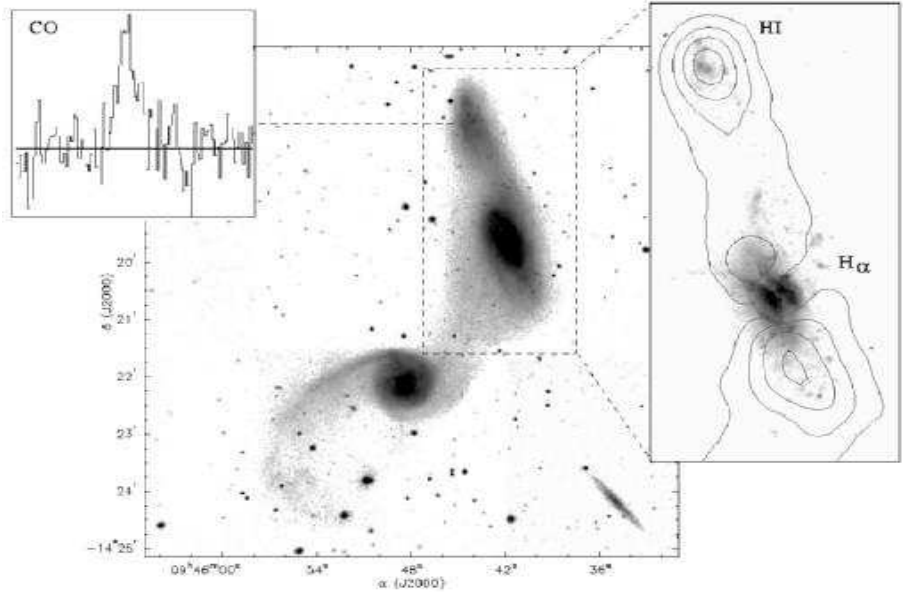,width=12cm,height=8cm}}
\caption[]{Outflows in the interacting system Arp 245. 
The inset to the right is a close-up of the spiral NGC 2992 and its
tidal tail. VLA HI contours are superimposed on an H$\alpha$ image.
This map exhibits a 30 kpc-long HI tail together with  10 kpc-long 
 H$\alpha$ filaments, escaping
 from the disk and which likely originate in an AGN.
Atomic, molecular (see the CO spectrum to the left) and ionized gas concentrate
 at the tip of the  tail fueling  a starburst.  Adapted from Duc et
al. (2000) and Braine et al. (2001).
}
\label{fig1}
\end{figure}

\section{Galactic material in the IGM/ICM}  
Besides hot gas and dark matter, the 
intracluster medium  contains matter more usually found in galaxies.
Star streams were discovered on deep optical images of clusters
(Gregg \& West, 1998).
Various surveys found numerous  planetary nebulae
 (Ford et al., 2001)  and red-giant stars (Ferguson et al., 1998)
floating between galaxies. From their numbers, it was extrapolated that the
intracluster stellar population may contribute between 5 and 
50\% to the total stellar mass in clusters.
 The neighborhood of colliding galaxies contains large quantities of 
atomic hydrogen. The percentage of extragalactic HI gas  observed in
emission at 21~cm  typically ranges between 50 and 90\% of the total HI
content of interacting systems. 
Even more surprisingly, 
extragalactic  molecular gas, as traced by the millimetre CO line, was
 detected in several groups,  in particular in Stephan's quintet
(see Fig.~2) where  we measured more than  $3 \times 10^{9}$ M$_{\odot}$
 of H$_2$.    

Where does such intergalactic material come from? A cosmological
origin  can be excluded. Indeed, optical
spectroscopy  indicates metallicities typical of galactic disks
that are inconsistent with primordial clouds.  
Therefore, this matter could either be the remnant of totally disrupted galaxies
 or expelled galactic material.

\begin{figure} 
\centerline{\psfig{file=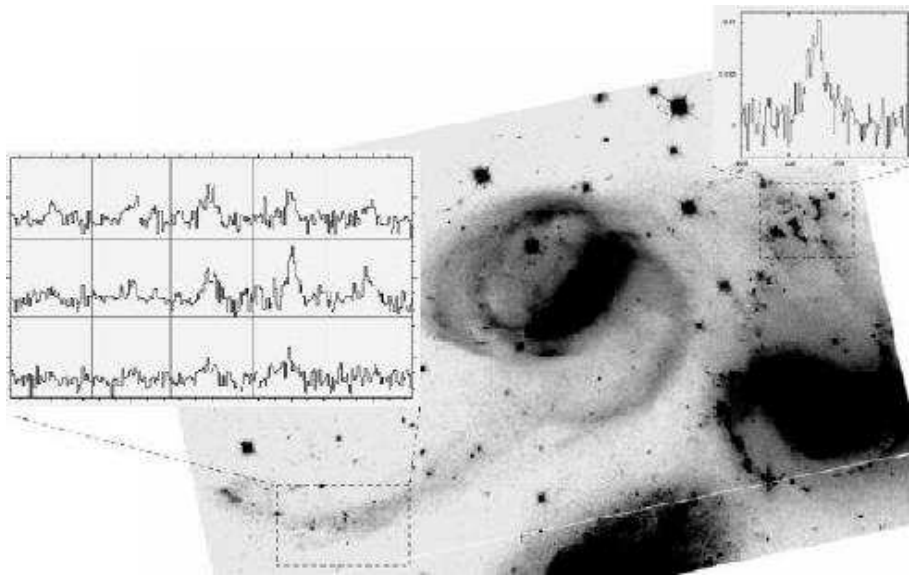,width=12cm}}
\caption[]{Intergalactic molecular clouds in the Hickson Compact Group HCG 92 (Stephan's quintet).
 CO(1-0) spectra
 obtained with the IRAM 30m antenna  are superimposed on an HST image of the group
 (Lisenfeld et al. 2002, in prep.)}
\label{fig2}
\end{figure}

\section{Fate of stripped material}  
The fate of galactic debris or ejecta will largely
depend on their nature, distance from the progenitors and on 
time scales.
First of all, simple gravitation will cause the ejecta to fall back, 
 eventually,  on to the parent galaxies. (Re)accretion 
 has since long been taken into account  in
semi-analytic models of galaxy evolution  and
 studied in detail using numerical simulations of galaxy collisions
 (Hibbard \& Mihos, 1995) or ram pressure stripping (Vollmer et al.,
 this volume). Time scales for reaccretion vary between several Myr and one
Hubble time depending on how far stripped material had been
ejected.
Lost material may be so diluted in the ICM that it
becomes barely visible. The stellar component will light up as a 
diffuse background.
The low-column density atomic hydrogen hitting the hot intracluster medium
will evaporate or become ionized, becoming invisible 
in the  21~cm line. 
 
Finally, part of the 'lost' material 
 is recycled directly within the intergalactic environment. This is the
origin of the so-called tidal dwarf galaxies (TDGs),
made out of tidal material pulled out from colliding galaxies.
These gas--rich, dynamically young objects are now commonly observed
near interacting systems (Weilbacher et al., 2000), in groups
 or clusters of galaxies. 
From our multiwavelength observations, we are now able to 
roughly analyze several processes pertaining to their formation. 
Using Fabry-Perot H$\alpha$  datacubes, we  identified
kinematically distinct entities decoupled from the streaming
motions which characterise the kinematics in the gaseous tidal tails.
 Their position-velocity diagrams 
(see  Fig.~3) show velocity gradients of typically 50~km~s$^{-1}$  over
 scales of a 1--5 kpc.
Spatially, such objects are
located at the peak of the HI column density. This is also precisely
where we detected abundant quantities of molecular gas which most
likely was produced in situ from the HI (Braine et al., 2001). 
 The scenario accounting for the 
formation of TDGs would hence involve an instability in the tidal HI,
its collapse and further transformation in to H$_2$ and the onset of
 star-formation. Unfortunately in order to confirm this, we still lack 
 the support from numerical simulations
that so far failed to produce tidal objects similar to those observed.  
Other pending questions are the global amount of material involved in 
such cosmic recycling  and the survival time of TDGs.

\begin{figure} 
\centerline{\psfig{file=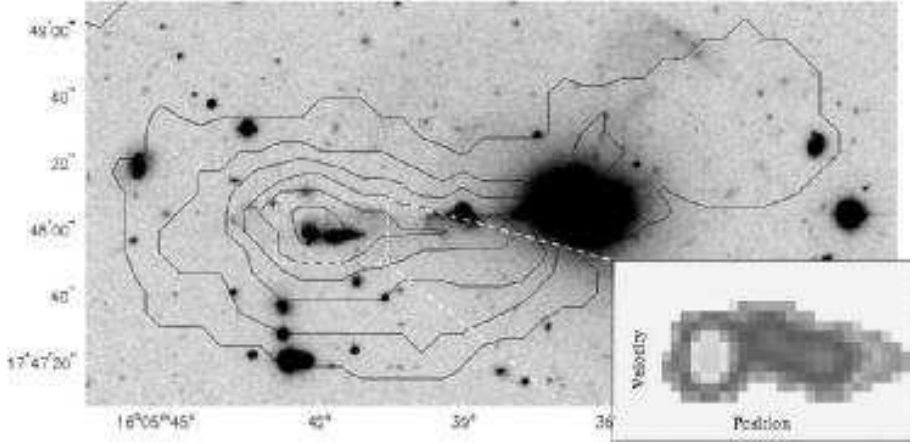,width=12cm}}
\caption[]{A tidal dwarf galaxy near the merger IC 1182, in the
Hercules cluster. VLA HI contours are superimposed on an optical
image of the system. The TDG -- the compact object to the left --
 is associated with a large HI condensation. The inset presents a
 position-velocity diagram in the ionized component obtained from
 Fabry-Perot observations at CFHT.
It exhibits a strong velocity gradient, which is
likely linked with the formation of the TDG.}
\label{fig3}
\end{figure}





\end{article}
\end{document}